\documentclass[review]{article}%

\usepackage{amsmath,amssymb} 
\usepackage{graphics,graphicx}
\usepackage[vmargin=2.5cm]{geometry}
\usepackage[colorlinks=true,urlcolor=black,linkcolor=black,citecolor=black]{hyperref}
\usepackage{color}
\usepackage{times}

\DeclareMathOperator*{\tra}{\,^{\mbox{\tiny T}}\!}
\newcommand{\bmat}{\left[\begin{matrix}}
\newcommand{\emat}{\end{matrix}\right]}

\begin{document}

  \title{Stop-and-go waves induced by correlated %(coloured) 
	noise in pedestrian models without inertia}

\author{Antoine Tordeux$^{1,}$\footnote{\href{mailto:tordeux@uni-wuppertal.de}{\texttt{tordeux@uni-wuppertal.de}}}~~ Andreas Schadschneider$^2$~ Sylvain Lassarre$^3$\\[2mm]
\small$^1$School of Mechanical Engineering and Safety Engineering, University of Wuppertal, Germany\\[1mm]
%{as@thp.uni-koeln.de}
\small$^2$Institute for Theoretical Physics, University of Cologne, Germany\\[1mm]
%{sylvain.lassarre@ifsttar.fr}
\small$^3$GRETTIA--COSYS, IFSTTAR, France}
\date{\footnotesize \today}
\maketitle

\begin{abstract}
  Stop-and-go waves are commonly observed in traffic and pedestrian
  flows.  In most traffic models they occur through a phase transition
  after fine tuning of parameters when the model has unstable
  homogeneous solutions.  Inertia effects are believed to play an
  important role in this mechanism. Here, we present a novel
  explanation for stop-and-go waves based on stochastic effects 
    in the absence of inertia.  The introduction of specific coloured
  noises in a stable microscopic first order model allows to describe
  realistic stop-and-go behaviour without requiring instabilities or
  phase transitions. We apply the approach to pedestrian single-file
  motion and compare simulation results to real pedestrian
  trajectories. Plausible values for the model parameters are
  discussed.\\[1mm]
  {\bf Keywords\,:}~~Pedestrian single-file motion; Stop-and-go dynamics; First-order
  microscopic models; Brownian noise; Simulation
\end{abstract}

\section{Introduction}
\label{sec:1}
%\begin{svgraybox}Existence of phase transitions for the stop-and-go dynamics in real traffic flow remains controversially\end{svgraybox}

%{\bf Updating of references needed!}

%{\bf maybe we can introduce "noise" as a "mild" form of stochasticity?}

As one of the characteristic collective phenomena in any kind of
traffic system, stop-and-go waves have attracted attention for a long
time now (see e.g.\
\cite{Chowdhury2000,Kernerbook,Treiber2013,Schadschneider2018,Chraibi2018}
for reviews).  Generically, congested flows show self-organisation in
the form of waves of slow and fast traffic instead of streaming
homogeneously.  This stop-and-go dynamics is not only observed in road
traffic, but also in bicycle and pedestrian streams \cite{Zhang2014},
both in real life and in controlled experiments. This is often called
spontaneous jam formation since the occurrence of the congestion can
not be explained by an (external) disturbance, e.g. due to the
infrastructure (bottlenecks) \cite{Sugiyama2008}. A thorough
understanding of such self-organisation phenomena phenomena will have
impact beyond the purely scientific aspects due to its relevance
e.g. for safety and comfort of transportation networks.

In order to study stop-and-go behaviour in traffic system often
continuous models based on non-linear differential systems are
used. Most models are based on second order systems and thus
  inertia. These models have homogeneous equilibrium (stationary)
solutions that can become unstable for certain values of the control
parameters. For the unstable cases, the solutions are non-homogeneous,
e.g. periodic or quasi-periodic. Stop-and-go waves can appear for fine
tuning of the parameters. This generic behaviour is found in many
microscopic, mesoscopic (kinetic) and macroscopic models based on
non-linear differential systems (see for instance
\cite{Bando1995,Helbing1998,Colombo2003}).  Typically these continuous
models are inertial second order systems based on relaxation
processes. When the inertia of the particles (vehicles,
pedestrians,...) exceeds a critical value
\cite{Bando1995,Muramatsu1999,Tomer2000}, stop-and-go behaviour occurs
that usually can be described by Korteweg--de Vries (KdV) or modified
KdV soliton equations.

Instabilities leading to phase transitions are observed in many
self-driven dynamical systems far from the equilibrium, e.g. in
physics, theoretical biology and social science
\cite{Ben-Jacob1994,Vicsek1995,Bussemaker1997,Buhl2006,Hermann2012}.
Empirical data and controlled experiments have provided evidence for
phase transitions and associated phenomena, like hysteresis or
capacity drop, mostly for vehicular traffic \cite{Kerner1997,Sugiyama2008}.
Currently, there is still some debate about e.g. the number of
phases and their characteristics \cite{Kernerbook,Treiber2010}.

For pedestrian dynamics the understanding of stop-and-go dynamics is
somewhat different.  To our knowledge, up to now, there is no
convincing empirical evidence for phase transitions and associated
instabilities in pedestrian flow.  Pedestrian dynamics shows no
pronounced inertia effects or mechanical delays since human capacity
allows nearly any speed variation at any time.  Nevertheless,
stop-and-go behaviour has been observed in pedestrian dynamics at
congested density levels \cite{Seyfried2010,Zhang2014}. Therefore, on
the theoretical level, most studies are based on ideas which are close
to that in vehicular traffic, i.e. a mechanism based on instability
and phase transitions
\cite{Portz2010,Moussaid2011,KUANG2012,Lemercier2012}.  
However, this is not very realistic for pedestrian dynamics where
  inertia effects play a much smaller role than in vehicular
  traffic. Inertia is also responsible for most artefacts like
  particle penetration, exceeding the desired velocity or unrealistic
  oscillatory motion that is sometimes observed in second order
  models. Therefore it seems much more natural to describe pedestrian
  flows by a first order approach.

In discrete stochastic models, e.g. cellular automata, the
origin of the stop-and-go waves is somewhat different
\cite{Barlovic1998,Knospe2000}. {By design,} in these models the
dynamics is very much determined by the stochasticity so that e.g. no
stable homogeneous solutions exist for any density. Traffic jams are
formed by fluctuations intrinsic to the dynamics. In this sense the
mechanism that we will propose here is much closer in spirit to
stochastic models than to (deterministic) models based on differential
systems.

Here we propose a novel explanation of stop-and-go phenomena in
pedestrian flows as a consequence of stochastic effects. Based on
statistical evidence for the existence of Brownian noise in pedestrian
speed time-series coming from single-file experiments, a microscopic, 
stochastic first-order longitudinal model is
proposed. The dynamics has only a minimal deterministic part for the motion.
%convection {\bf (ist diese Terminologie allgemein bekannt?)}. 
In addition, a relaxation process for the noise is introduced.  Based on
computer simulations we show that the model allows to describe
realistic pedestrian stop-and-go dynamics without instability and fine
tuning of the parameters.

%The definition of the model and some its properties are presented in Sect.~\ref{sec:2}.  Some numerical results of the model on a ring and comparison to real pedestrian data are proposed in Sect.~\ref{sec:3}, while discussion concluded the article in Sect.~\ref{sec:4}.

%%%%%%%%%%%%%%%%%%%%%%%%%%%%%%%%%%%%%%%%%%%%%%%%%%%%%%%%%%%%%%%%%%%%%%%%%%%

\section{Definition of the stochastic model}
\label{sec:2}

Stochastic effects can have various roles in the dynamics of
self-driven systems \cite{Haenggi2007}.  The introduction of white
noise in models tends to increase the disorder in the system
\cite{Vicsek1995} or prevents self-organisation \cite{Helbing2000}.
On the other hand, coloured noises can affect the dynamics and
generated complex patterns \cite{Arnold1978,Castro1995}. For traffic
systems it is interesting to not that coloured noise has been
observed in human response \cite{Gilden1995,Zgonnikov2014}.
%$1/f$ or pink noises for instance are observed in many complex
%systems are characteristic of self-organized criticality
%\cite{Bak1987}.

Human behaviour in traffic results from complex human cognition. It is
intrinsically stochastic in the sense that the deterministic modelling
of the human cognition based on the states and interactions of up to
10$^\text{11}$ neurons \cite{Williams1988} is not possible.  The
behaviour in traffic is furthermore influenced by multiple factors,
e.g.\ experience, culture, environment, psychology, etc. as well as
random external events that can not be fully captured by any model.
Based on the experience from the field of Statistical Physics this
lack of knowledge in complex systems can usually be captured well by
introducing stochasticity into the dynamics. Indeed, stochastic
effects are not only an essential part of cellular automata based
approaches but have also been introduced in many traffic models based
on differential equations (usually in the form of noise), 
%and notion of noise are the main emphasis of many pedestrian or road
%traffic modelling approaches (see,
e.g. as white noise \cite{Helbing1995,Tomer2000}, pink-noise
\cite{Takayasu1993}, action-points \cite{Wagner2006}, or inaccuracies
and risk-taking behaviour \cite{TREIBER2006,Hamdar2015}.  Yet in
contrast to stochastic cellular automata for which the rule is
intrinsically stochastic, adding a noise in differential systems is a
mild form of stochasticity.

\subsection{Empirical evidence for Brownian noise}
\label{Subsec-Brown}

\begin{figure}[!ht]\vspace{5mm}
\begin{center}
\includegraphics[scale=1.1]{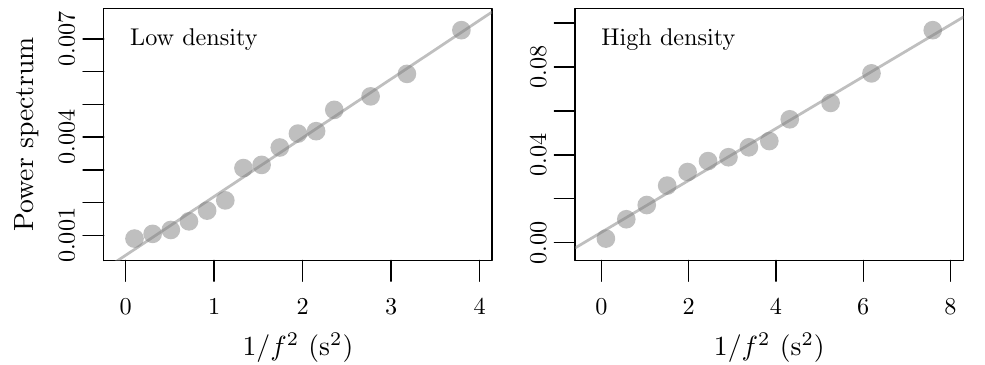}
\caption{Periodogram power spectrum estimate for the speed time-series
  of pedestrians at low and high density levels.  The power spectrum
  is roughly proportional to the inverse of square frequency $1/f^2$.
  This is a typical characteristic of Brownian noise.}
\label{fig:1}       % Give a unique label
\end{center}
\end{figure}

Fig.~\ref{fig:1} shows a typical time series of pedestrian speeds for
trajectories coming from single-file experiments (see
\cite{Portz2010,data,Tordeux2016} for details on the data).  The power
spectral density (PSD) is found to be proportional to $1/f^2$ where
$f$ is the noise frequency.  This frequency dependence is the hallmark
of Brownian noise has a PSD proportional to the inverse of the square
noise frequency $1/f^2$,  
%A characteristic linear tendency is observed
independent of the density.  Such a noise with exponentially
decreasing time-correlation function can be described by using the
Ornstein-Uhlenbeck process (see e.g.\ \cite{lindgren2013}).  Note that
comparable tendencies are observed for traffic flow as well
\cite{Hamdar2015}.

%%%%%%%%%%%%%%%%%%%%
%%%%%%%%%%%%%%%%%%%%

\subsection{Model definition}

%{\bf (The description seems to refer to single-file experiments! If true, that should be mentioned!)}

%The experimental data described in Sec.~\ref{Subsec-Brown} has been obtained in experiments of single-file motion in a corridor with periodic boundary conditions {\bf (????)}. 
In the following we
therefore introduce a continuous stochastic model to describe one-dimensional pedestrian
motion in single-file experiments.
We denote the curvilinear position of pedestrian $k$ at time $t$ by
$x_k(t)$. Pedestrian $k+1$ is the predecessor of $k$. The motion of
the pedestrian $k$ is then described by the Langevin equation
\cite{Tordeux2016}
\begin{equation}
\left\{\begin{array}{lcl}
  \mbox d x_k(t)&=&V\big(x_{k+1}(t)-x_k(t)\big)\,
  \mbox dt+\varepsilon_k(t)\,
  \mbox dt,\\[.5mm]
  \mbox d\varepsilon_k(t)&=&-\frac1\beta\varepsilon_k(t)\,
  \mbox dt+\alpha\,
  \mbox d W_k(t).
\end{array}\right.
\label{mod}
\end{equation} 
where $V:s\mapsto V(s)$ is a differentiable and non-decreasing
\emph{optimal velocity} (OV) function for the convection
\cite{Bando1995}. The noise $\varepsilon_k(t)$ is described by an
Ornstein-Uhlenbeck stochastic process \cite{OU}.  

For simplicity, the OV function is chosen as an affine function
\begin{equation}
V(s)=\frac1T(s-\ell)
\end{equation} 
in the following, where $T$ is the time gap between the agents 
and $\ell$ their size. This form is not realistic for very
small densities since it is not bounded by some maximal velocity.
However, here we are only interested in the congested regime, i.e.
intermediate and high densities so that this problem is irrelevant
since our systems are one-dimensional with periodic boundary
conditions. 
The quantities $(\alpha,\beta)$ in (\ref{mod}) are parameters related
to the noise: $\alpha$ is the volatility and $\beta$ the noise
relaxation time.  Finally, $W_k(t)$ is a Wiener process.

The model can be considered as a special stochastic variant of the
Full Velocity Difference model \cite{Jiang2001}.  Writing
$\dot x_k=\mbox d x_k/\mbox d t$,
$\ddot x_k=\mbox d^2 x_k/\mbox d^2 t$ and $\mbox d W_k/\mbox dt=\xi_k$
as a white noise, one gets from Eq.~(\ref{mod}) the second order
system% $\varepsilon_k=V\big(x_{k+1}-x_k\big)-\dot x_k$ and
\begin{equation}
\ddot x_k=\big[V(x_{k+1}-x_k)-\dot
x_k\big]/\beta+V'(x_{k+1}-x_k)(\dot x_{k+1}-\dot x_k)+\alpha\xi_k.
\label{fvdm}
\end{equation}
This is a noisy version of the Full Velocity Difference model
\cite{Jiang2001} for which the relaxation time for the speed
difference is the derivative of the optimal velocity function.

At this point we want to emphasize a characteristic property of the
proposed model. The convection part (first equality in (\ref{mod})) is
of first order, while the noise operates at second order (second
equation in (\ref{mod})). %In contrast to the classical approaches, where generally the convection part is relaxed, relaxation here takes place due to the noise. {\em (The convection part has also a relaxation towards the optimal velocity...)}
The first order nature of the convection part reflects the assumption
that for pedestrian motion inertia effects are less relevant than for
vehicular traffic. Instead it is often assumed that pedestrians can
stop or accelerate immediately which is naturally described by
a first order equation without a mass term.

%%%%%%%%%%%%%%%%%%%%%%%%%%%%%%%%%%%%%%%%%%%%%%%%%%%%%%%%%%%%%%%%%%%%%%%%%%%

\section{Stability analysis}

\begin{figure}[!ht]\vspace{5mm}
\begin{center}
\includegraphics[scale=1.1]{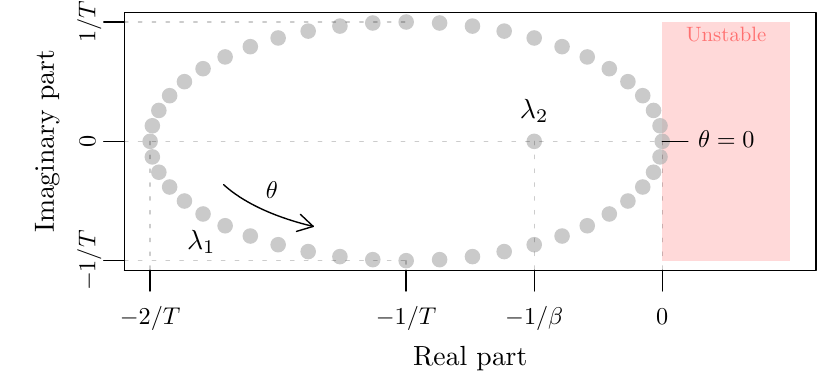}
\caption{Eigenvalues for the matrix of the linear system for
  $n=50$. The homogeneous solution is stable when all the eigenvalues
  have strictly negative real part excepted the one for $\theta=0$,
  which is null.}
\label{fig:2}       % Give a unique label
\end{center}
\end{figure}

We now consider one-dimensional motion of $n$ agents on a line of
length $L$ with periodic boundary conditions. The dynamics defined in
Eq.~(\ref{mod}) can be written as a system of equations,
\begin{equation}
\mbox d \eta(t)=(A\eta(t)+a)\,\mbox dt+b\,\mbox d w(t)
\label{syst}\end{equation}
with 
\begin{eqnarray}
\eta(t)&=&\tra\left(x_1(t),\varepsilon_1(t),\ldots,x_n(t),
\varepsilon_n(t)\right), \nonumber\\
a&=&-\frac{\ell}T\tra(1,0,\ldots,1,0), \\
b&=&\alpha\tra(0,1,\ldots,0,1) \nonumber
\end{eqnarray}
and
\begin{equation}
%\mbox{$A=$\small$\bmat R&S&\\[-2.4mm]&\ddots&\\[-2.4mm]S
%&&R\emat$\normalsize$\qquad$with$\quad$ 
%$R=$\small$\bmat -1/T&1\\0&-1/\beta\emat$
%\normalsize$\quad$and$\quad$$S=$\small$\bmat 1/T&0\\0&0\emat$}.
A=\bmat R&S&\\[-2.4mm]&\ddots&\\[-2.4mm]S&&R\emat
\quad \text{with\ \ \ }
R=\bmat -1/T&1\\0&-1/\beta\emat, \quad S=\bmat 1/T&0\\0&0\emat\,.
\end{equation}
$A$ is a real $2n\times 2n$ matrix, while $\eta(t)$, $a$, $b$
are real $2n$-component vectors. $w(t)$ is $2n$-vector
composed of independent Wiener processes.
%\in\mathbb R^{2n\times2n}$ while $\eta(t)$, $a$, $b\in\mathbb R^{2n}$. 
%The $2n$-vector $w(t)$ is composed of independent Wiener processes.  
Such a linear stochastic process is Markovian. %and ergodic
It has a normal distribution with expectation $m(t)$ and
variance/covariance matrix $C(t)$ such that, by using the Fokker-Planck
equation (Kolmogorov forward equation)%{\bf (How, more details? or reference?)},
\begin{equation}
\dot m(t)=Am(t)+a\qquad\mbox{and}\qquad\dot C(t)=AC(t)+(C(t))\!\tra A
+\mbox{diag}(b)\quad
\end{equation} 
with $m(0)=\eta_0$ and $C(0)=0$.  Asymptotically the expectation value
$m(t)$ is given by the homogeneous solution for which
$x_{k+1}(t)-x_k(t)=L/n$ and $\varepsilon_k(t)=0$ (for all $k$ and 
$t$ and by taking $x_{k+1}-x_k=x_{1}-x_{n}$ for $k=n$). 
%$L$ being the length of the system. 
The matrix $A$ is circulant with blocks of size $2\times 2$.  Its
eigenvalues are those of $R+S\iota_\theta$, with
$\iota_\theta=e^{i\theta}$ where $\theta=2\pi k/N\in(0,2\pi)$
($k=0,\ldots,n-1$) is the $n$-th root of unity.  
The eigenvalues are then the solutions of the characteristic equation
\begin{equation}
%\textstyle
\left[\lambda+\frac1T\big(1-e^{i\theta}\big)\right]
\big[\lambda+1/\beta\big]=0\,.
\end{equation}
The solutions are $\lambda_1=-\frac1T\big(1-e^{i\theta}\big)$ and
$\lambda_2=-1/\beta$.  They have strictly negative real parts
$\mbox{Re}(\lambda_1)=-\frac1 T(1-\cos(\theta))$ and
$\mbox{Re}(\lambda_2)=-1/\beta$ for any $\theta\in(0,2\pi)$, $T>0$ and
$\beta>0$. Therefore the homogeneous solutions are linearly stable for
the system~(\ref{syst}) for any positive value of the parameters.  The
least stable configuration is the one with maximal period for which
$\theta\rightarrow0$ (see Fig.~\ref{fig:2}).

%%%%%%%%%%%%%%%
%%%%%%%%%%%%%%%

Note that the general stability conditions for 
2nd order models given for instance in \cite[Chap.~15]{Treiber2013} are
\begin{equation}
a>0,\qquad b+|c|<0,\qquad b^2-c^2>2a\,,
\label{CS}
\end{equation}
where $a$, $b$ and $c$ are the partial derivatives of the model
according to the distance spacing, the speed of the considered
vehicle, and the speed of its predecessor, respectively.  
For the 2nd order formulation of the model given in Eq.~(\ref{fvdm})
this implies
\begin{equation}
a=\frac1{T\beta},\qquad b=-\frac1T-\frac1\beta,\qquad c=\frac1T \,.
\end{equation}
It is easy to check that all the three conditions in Eq.~(\ref{CS}) 
hold, i.e.\ the homogeneous solution are deterministically stable, as
soon as $T>0$ and $\beta>0$. This confirms the results obtained above.

%%%%%%%%%%%%%%%%%%%%%%%%%%%%%%%%%%%%%%%%%%%%%%%%%%%%%%%%%%%%%%%%%%%%%%%%%%%

\section{Numerical experiments}
\label{sec:3}

We have simulated the system (\ref{syst}) using an explicit
Euler-Maruyama numerical scheme with time step $\delta t=0.01$~s.  The
other parameter have been chosen as $T=1$~s, $\ell=0.3$~m,
$\alpha=0.1$~ms$^{-3/2}$ and $\beta=5$~s which is close to estimates
for pedestrian flow \cite{Tordeux2016}.  Corresponding to the
experimental situation, the system length is $L=25$~m with periodic
boundary conditions.

\begin{figure}[!ht]\vspace{2mm}
\begin{center}
\includegraphics[width=.9\textwidth]{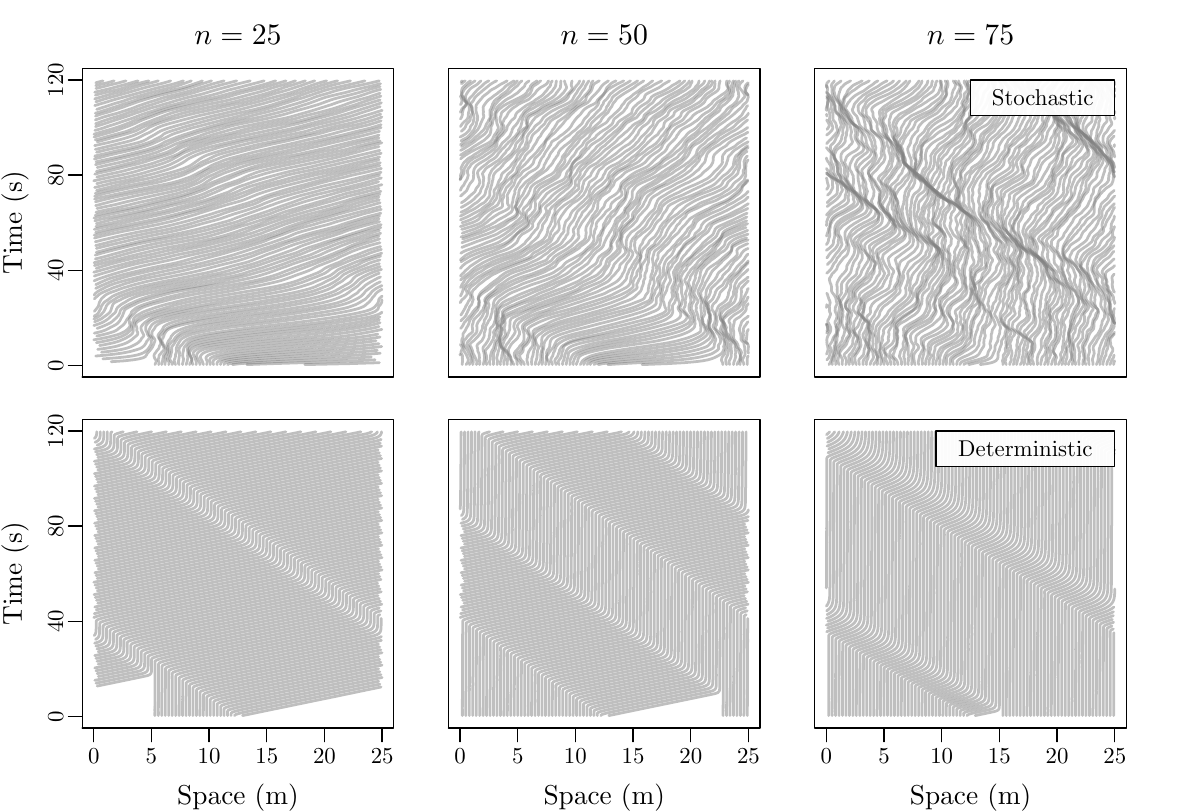}
\caption{Simulated trajectories for different densities ($n=25,50,75$
  pedestrians, $L=25$~m) and jam initial conditions.  Top panels:
  Stochastic model defined by (\ref{syst}). Bottom panels: Unstable
  deterministic model from \cite{Tordeux2014}. Parameter values are
  $T=1$~s and $\ell=0.3$~m for the affine optimal velocity of both
  models, $\alpha=0.1$~ms$^{-3/2}$ and $\beta=5$~s for the noise of
  the stochastic model and $T^r=0.7$~s for the reaction time parameter
  of the deterministic model.}
\label{fig:3}
\end{center}
\end{figure}

Simulations have been performed for the model defined by
Eq.~(\ref{syst}) and, for comparison, the unstable deterministic
optimal velocity model with two predecessors in interaction introduced in \cite{Tordeux2014}
\begin{equation}
\dot x_k=V\big(x_{k+1}-x_k-T^r\big[V(x_{k+2}-x_{k+1})-V(x_{k+1}-x_k)\big]\big)
\end{equation}
Here the optimal velocity function and value is the same as in the
stochastic model, i.e.\ $V(s)=\frac1T(s-\ell)$ with also $T=1$~s,
$\ell=0.3$~m, while the reaction time parameter is set to $T^r=0.7$~s
in order to describe unstable homogeneous configuration.
% (GIVE MODEL DEFINITION?).  
Starting from a jam initial condition the dynamics of
$n=25$, $50$ and $75$ pedestrians has been simulated for both models.
Fig.~\ref{fig:3} shows typical trajectories for the first two minutes.

\begin{figure}[!ht]
\begin{center}
\includegraphics[scale=1.1]{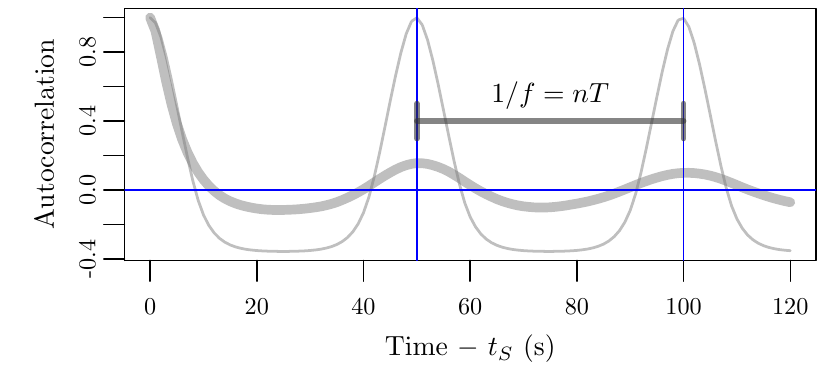}
\caption{Stationary mean time-correlation function of the distance
  spacing for the stochastic (thick line) and the deterministic
  model (thin line).  For stop-and-go waves, the same period
  $1/f=nT$ is observed.  The system can be considered 
  stationary after $t_S=\,$2$\cdot$10$^\text5$~s. }
\label{fig:4}  
\end{center}
\end{figure}

Fig.~\ref{fig:4} shows the mean autocorrelation functions for the
distance spacing for large simulation times $t>t_S$, with
$t_S\,$2$\cdot$10$^\text5$~s, where the system can be considered to be
in the stationary state.  The peaks of the autocorrelations match for
both models, indicating identical frequencies of the stop-and-go
waves.  A wave propagates backwards in the system at speed $c=-\ell/T$
while vehicles travel forward with average speed $v=(L/n-\ell)/T$,
where is $T$ the time gap parameter of the optimal velocity function.
In agreement with the LWR theory for traffic flow
\cite{Richards1956,Lighthill1955}, the wave period is $L/(v-c)=nT$.

\begin{figure}[!ht]\vspace{5mm}
\begin{center}
\includegraphics[width=.9\textwidth]{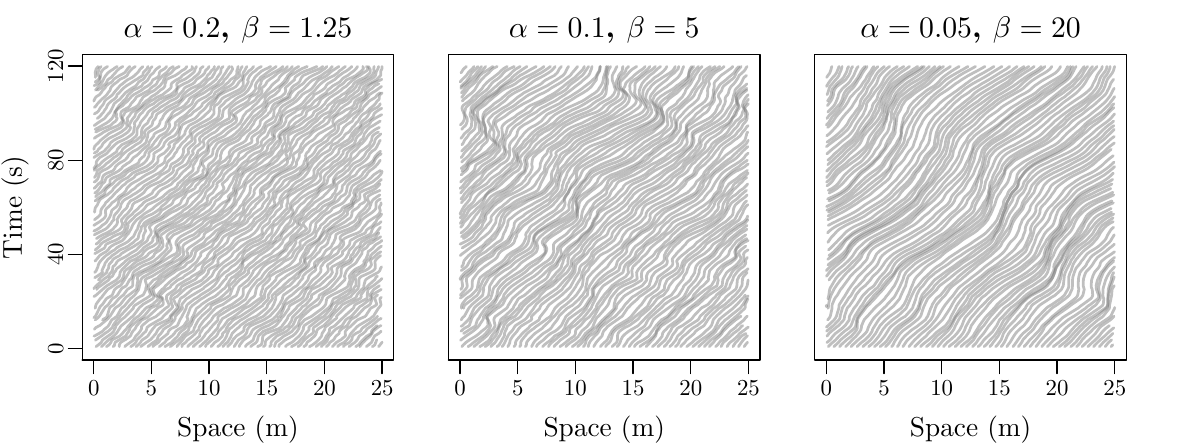}
\caption{Simulated trajectories of $n=50$ agents for different values
  of the noise parameters (units: $\alpha$ in ms$^{-3/2}$, $\beta$ in
  s).  The initial configuration is homogeneous.}
\label{fig:5} 
\end{center}
\end{figure}

\begin{figure}[!ht]\vspace{0mm}
\begin{center}
\includegraphics[scale=1.1]{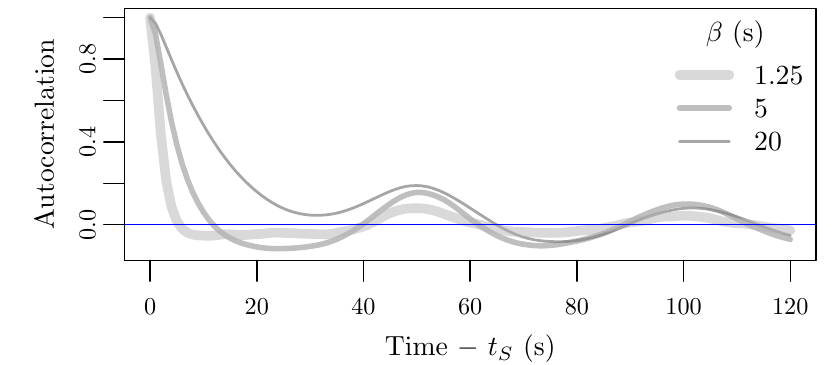}
\caption{Mean temporal correlation function of the distance spacing in
  the stationary state of (\ref{syst}) for different values of noise
  parameter $\beta$. $\alpha$ is chosen such that the noise amplitude
  is the same in all cases. The noise parameters do not influence the 
  frequency of the waves which only depends on $n$ and $T$.
%The amplitude increases when the relaxation time $\beta$ is high and
%inversely. 
%The simulation delay time is $t_S=\,$2$\cdot$10$^\text5$~s.
}
\label{fig:6}
\end{center}
\end{figure}

We have also determined the dependence of the results on the noise
parameters $\alpha$ and $\beta$.  Fig.~\ref{fig:5} shows trajectories
of $50$ agents for $\alpha=0.05$, $0.1$ and $0.2$~ms$^{-3/2}$. The
values of $\beta$ are chosen such that the amplitude of the noise
$\sigma=\alpha\sqrt{\beta/2}$ is constant, i.e. $\beta=1.25$, $5$ and
$20$~s, respectively.  For small relaxation times $\beta$, the noise
tends to be white and unstable waves emerge locally and disappear (see
Fig.~\ref{fig:5}, left panel).  For large $\beta$, on the other hand,
the noise autocorrelation is high.  In this case stable waves with
large amplitude occur (Fig.~\ref{fig:5}, right panel).
Not that the noise parameters influence only the amplitude of the
time-correlation function, but not the frequency that only depends on
the parameters $n$ and $T$ (see Fig.~\ref{fig:6}).

\begin{figure}[!ht]\vspace{5mm}
\begin{center}
\includegraphics[width=.9\textwidth]{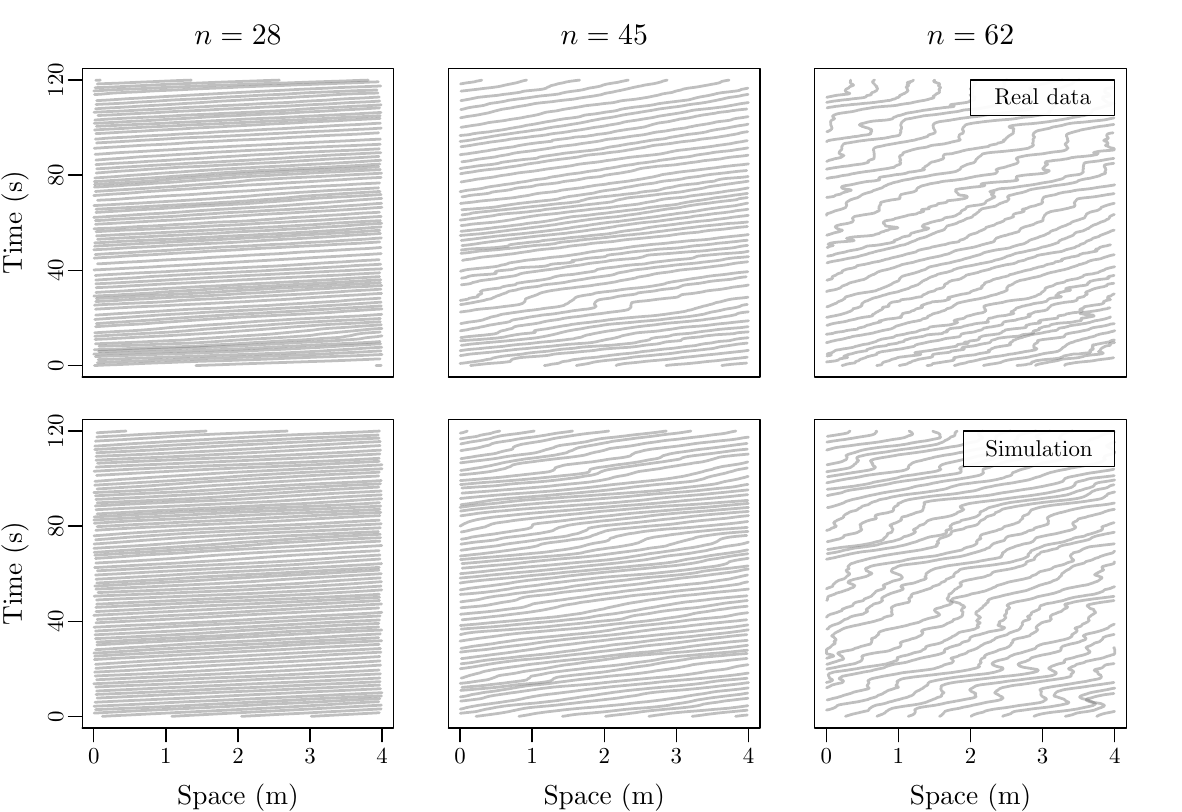}
\caption{Empirical (top panels) and simulated (bottom panels) trajectories
  for different densities.  The initial configuration is
  homogeneous both in the experiments and the simulations.}
\label{fig:7}       
\end{center}
\end{figure}

In Fig.~\ref{fig:7} empirical trajectories from experiments with $28$,
$45$ and $62$ participants \cite{Tordeux2016} are compared with
simulations of the stochastic model.  
%Corresponding mean time-correlation functions are drawn in Fig.~\ref{fig:6}.  
The data show a good agreement. Homogeneous free flow states are
observed for $n=28$ agents in both cases, while stop-and-go waves
appear in the semi-congested ($n=45$) and congested ($n=62$) states
in both data sets.
%The autocorrelations globally match.

%\begin{figure}[!ht]\vspace{-7mm}\includegraphics{figure6.pdf} \caption{Time-correlation function of the distance spacing for different densities.  The simulations are in good agreement with the empirical data.}\label{fig:6}\end{figure}

%%%%%%%%%%%%%%%%%%%%%%%%%%%%%%%%%%%%%%%%%%%%%%%%%%%%%%%%%%%%%%%%%%%%%%%%%%%

\section{Discussion}
\label{sec:4}

\begin{figure}[!ht]\vspace{5mm}
\begin{center}
\includegraphics[scale=1.1]{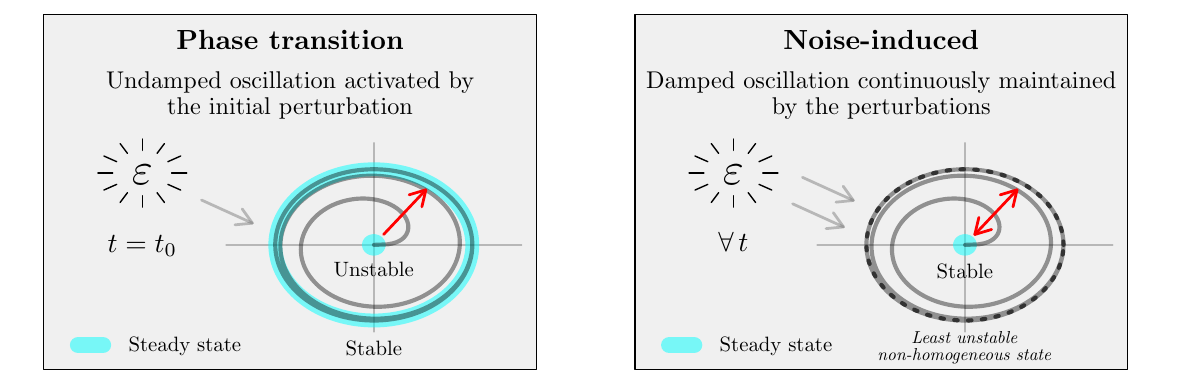}
\caption{Illustrative scheme for the modelling of stop-and-go dynamics
  with phase transition in the periodic solution (left panel) and the
  noise-induced oscillating behaviour (right panel).}
\label{fig:8}    
\end{center}
\end{figure}

We have presented an alternative explanation for the occurrence of
stop-and-go phenomena in pedestrian flows. In contrast to previous
explanations, the formation of stop-and-go waves here is the
consequence of coloured noise in the dynamics of the pedestrian speeds
which has been observed in empirical data for single-file motion.
This correlated noise can provide perturbations that lead to
oscillations in the system, especially when the system is poorly
damped.
%In this stochastic approach, 
%at its deterministic characteristic frequency
%are due to the perturbations provided by the noise.  
%Such a mechanism qualitatively describes stop-and-go waves  
This new mechanism differs that in classical deterministic traffic
models with inertia. Here stop-and-go waves occur as a consequence of
the instability of the homogeneous configuration. In such a
  situation s single small perturbation $\varepsilon$ is sufficient to
  drive the system via a phase transition into a state with periodical
  dynamics (see Fig.~\ref{fig:8}, left).  In the noise-induced
  mechanism proposed here the correlated noise can 'kick' the system
  out of its stable homogeneous state into a non-homogeneous state in
  which a damped oscillation is continuously maintained by the
  perturbations from the noise (see Fig.~\ref{fig:8}, right).

%We have identified two mechanisms based on relaxation processes that
%are relevant for the description of stop-and-go waves.  
The two mechanisms differ also in the relevant relaxation
  processes. In the novel stochastic approach, the relaxation time is
related to the noise and is estimated to approximately 5~s
\cite{Tordeux2016}.  The parameter corresponds to the mean time period
of the stochastic deviations from the phenomenological equilibrium
state.  Such a time can be large, especially for small deviations and
large spacings. In contrast, in the classical inertial approaches, the
relaxation time is interpreted as the driver/pedestrian reaction time
and is estimated to around 0.5 to 1~s.  Technically, such a parameter
can not exceed the physical time gap between the agents (around 1 to
2~s) without generating unrealistic behaviour (e.g. collisions) and has
to be set carefully for these models.

We believe that the new mechanism based on a first order convection
equation is specially relevant for pedestrian dynamics. The motion of
pedestrians is believed to be much less influenced by inertia effects
than the motion of vehicles and is thus much more effect by
  noise, especially by correlated noises.  Due to the limited inertia
  effects a description based on a first order model is much more
  natural for pedestrian dynamics. This would also avoid many of the
  artefacts observed in force-based models (see
  e.g. \cite{ChraibiETNSS15} and references therein) which are often a
  consequence of strong inertia effects.  In future studies we expect
an even better agreement with empirical data when more realistic
optimal velocity functions are used in the model. Furthermore, extensions 
of the model should be carried out to describe the motion of pedestrians 
in two dimensions, including models for the direction and the definition 
and meaning of time gap in 2D.

%%%%%%%%%%%%%%%%%%%%%%%%%%%%%%%%%%%%
%%%%%%%%%%%%%%%%%%%%%%%%%%%%%%%%%%%%

\section*{Conflict of interest}
The authors do not have any conflict of interest with other entities or researchers.

\section*{Acknowledgement}
%Please see the title page.
The authors thank Prof.\ Michel Roussignol for his help in the formulation of the model. 
Financial support by the German Science Foundation under grant SCHA 636/9-1 is gratefully acknowledged.

%\section*{References}

%\bibliographystyle{unsrt}
%\bibliography{bibli}

\end{document}